\documentclass[
showpacs,preprintnumbers,amsmath,amssymb]{revtex4}
\usepackage{amsmath}
\usepackage{graphicx}
\usepackage{subfig}
\usepackage{hyperref}
\usepackage{amssymb}
\usepackage[english]{babel}
\usepackage{epsfig}
\usepackage{wasysym}
\usepackage{graphicx}
\usepackage{indentfirst}
\usepackage{amsmath}
\usepackage{amsfonts}
\usepackage{amssymb}
\usepackage{enumerate}

\begin{document}
\def\be{\begin{equation}}
\def\ee{\end{equation}}
\def\bea{\begin{eqnarray}}
\def\eea{\end{eqnarray}}


\title{A Type of Levi-Civita's Solution in Modified Gauss-Bonnet Gravity}
\author{M. E. Rodrigues $^{(a,e,f)}$\footnote{E-mail
address: esialg@gmail.com}, M. J. S. Houndjo $^{(b)(c)}$\footnote{E-mail address:
sthoundjo@yahoo.fr}, D. Momeni $^{(d)}$\footnote{E-mail: d.momeni@yahoo.com} and R. Myrzakulov $^{(d)}$\footnote{E-mail address: 
rmyrzakulov@csufresno.edu} }
\medskip
\affiliation{$^{(a)}$Universidade Federal do Esp\'{\i}rito Santo \\
Centro de Ci\^{e}ncias
Exatas - Departamento de F\'{\i}sica\\
Av. Fernando Ferrari s/n - Campus de Goiabeiras - CEP29075-910 -
Vit\'{o}ria/ES, Brazil}
\affiliation{$^{(b)}$ Departamento de Ci\^{e}ncias Naturais - CEUNES\\
Universidade Federal do Esp\'irito Santo - 
CEP 29933-415 - S\~ao Mateus - ES, Brazil}
\affiliation{$^{(c)}$Institut de Math\'{e}matiques et de Sciences Physiques (IMSP) - 01 BP 613 Porto-Novo, B\'{e}nin}
\affiliation{$^{(d)}$Eurasian International Center for Theoretical Physics - Eurasian National University,\\ Astana 010008, Kazakhstan}
\affiliation{$^{(e)}$Faculdade de Ci\^{e}ncias Exatas e Tecnologia, Universidade Federal do Par\'{a} - Campus Universit\'ario de Abaetetuba, CEP 68440-000, Abaetetuba, Par\'{a}, Brazil}
\affiliation{$^{(f)}$Faculdade de F\'{\i}sica, Universidade Federal do Par\'{a}, 66075-110, Bel\'em, Par\'{a}, Brazil}

\begin{abstract}
In this letter we obtain an exact solution for cylindrically symmetric modified Gauss Bonnet gravity. This metric is a generalization of the vacuum solution of the Levi-Civita in the general relativity. It describes an isotropic perfect fluid one parameter family of the gravitational configurations which can be interpreted as the exterior metric of a cosmic string. By setting the Gauss-Bonnet coupling parameter to zero, we recover the vacuum solution in the Einstein gravity as well. 
\end{abstract}
\maketitle 

\section{Introduction}

Einstein field equations are the simplest toy models for description  the gravity as a geometrical object, due to the Mach principle in terms of the classical gauge theory. The equivalence principle and the tensor form of the physical rules give us more freedom to introduce another models, which so called as the modified gravity models \cite{review}. To finding an exact solution for any types of the gravity models we need to solve a system of highly non linear differential or partial equations. As we know the solution of a non linear system is not unique and so by any assumption we can find another solution. Also, here there is no gauge freedom to connect the different kinds of the solutions. Just some generating techniques have been proposed as Buchdahl transformation \cite{buchdahl}, Ehlers transformations \cite{ehler}, and recently the fluid-gravity duality conjecture from the string theory opened a new window to the exact solutions \cite{fluid.gravity}. The last approach is more advanced one and potentially is very rich to extension. 
\par
Form other point of the view, symmetry is a basic key object for description of the physical systems appropriately. By fixing the symmetry of the metric in the gravitational models, we archive a tool for simplification of the systems and reduction of the numbers of the independent differential equations. In spite of the Newtonian mechanics, in the gravitational theories, usually the symmetry of the source can not be addressed directly in the symmetry of the metric of the manifold. For example a rod like source also can produced the spherically symmetric static configurations as well as a member of the Papapetrou class. The spherical symmetry is trivial and first one. By applying this symmetry we can obtain a family of black objects in any dimension $d$ also when the static configuration is formed in the $d>4$. The horizon of these black objects can be in any form from $S^2$ to $S^2\times S^p$ (in $d>4$).
\par 
The cylindrical symmetry is the next symmetry which can be considered properly. Most of the black objects in the higher dimensions and with additional fields like $U(1)$ electromagnetism have a class of the exact solutions with this posses this kind of the symmetry. In fact, this symmetry is a subclass of the axial symmetries. Different kinds of the exact solutions with cylindrical symmetry have been discussed in the literature, in f(R) \cite{f(R)} and also in forth order  Weyl gravity \cite{weyl} and also
in braneworld gravity\cite{brane}, Einstein-Maxwell \cite{EM} and with superconducting cosmic strings \cite{super}. In five dimensions, also you can have a cosmic string structure which is a vacuum solution to five dimensional field equations \cite{gb}. In this letter we want to find exact solutions, in $d=4$, for a modified gravity with both curvature ``R'' and the Gauss-Bonnet invariant term ``G'' in cylindrical symmetry. 
\par
We derived the field equations for a typical Lagrangian density in the form $\mathcal{L}=R+f(G)$ where in it, $f(G)$ is an arbitrary function of the Gauss-Bonnet topological invariance. By assuming a viable forms of the $f(G)\sim G^n$ we obtain an exact solutions which it satisfies all the field equations. The exact solution correctly recovers the cylindrical vacuum solution of the Levi-Civita appropriately and we found the expressions for the pressure and the energy density values. So, our exact solution is a natural extension of the vacuum cylindrically symmetric solution in the modified Gauss Bonnet gravity in $d=4$.

\section{Formalism of $f(G)$ gravity and equations of motion within cylindrical metric}

\hspace{-0,45cm}Let us consider expose a brief review of the field equations in $R+f(G)$ theory of gravity \cite{nojiri1} using the action given by
\begin{eqnarray}
S=\int d^4x\sqrt{-g}\left[\frac{R}{2\kappa^2}+f(G)\right]+S_m\,\,,\label{ratbay1}
\end{eqnarray}
where $R$ is the Ricci scalar, $f(G)$ is a general differential function of the Gauss-Bonnet invariant $G$.  $S_m$ is the matter action that from which the energy momentum tensor $T_{\mu\nu}$ will be derived, and $\kappa^2=8\pi G_N$, where $G_N$ is the Newtonian gravitational constant. By varying the action (\ref{ratbay1}) with respect to
the metric $g_{\mu\nu}$, one gets the following equation
\begin{eqnarray}
R_{\mu\nu}-\frac{1}{2}Rg_{\mu\nu}+8\Big[R_{\mu\rho\nu\sigma}+R_{\rho\nu}g_{\sigma\mu}
-R_{\rho\sigma}g_{\nu\mu}-R_{\mu\nu}g_{\sigma\rho}+R_{\mu\sigma}g_{\nu\rho}\nonumber\\
+\frac{R}{2}\left(g_{\mu\nu}g_{\sigma\rho}-g_{\mu\sigma}g_{\nu\rho}\right)\Big]\nabla^{\rho}\nabla^{\sigma}f_{G}+\left(Gf_{G}-f\right)g_{\mu\nu}=\kappa^2 T_{\mu\nu}\,\,,\label{ratbay2}
\end{eqnarray}
where $f_{G}$ denotes the derivative of the function $f$ with respect to $G$, defined as $G=R^2-R_{\mu\nu}R^{\mu\nu}+R_{\mu\nu\lambda\sigma}R^{\mu\nu\lambda\sigma}$, $R_{\mu\nu}$ and $R_{\mu\nu\lambda\sigma}$ being the Ricci tensor and Riemann tensors, respectively. The conventions used in this work are $(+---)$ for the signature, $\nabla_{\mu}V_{\nu}=\partial_{\mu}V_{\nu}-\Gamma_{\mu\nu}^{\lambda}V_{\lambda}$ and $R^{\sigma}_{\;\mu\nu\rho}=\partial_{\nu}\Gamma^{\sigma}_{\mu\rho}-\partial_{\rho}
\Gamma^{\sigma}_{\mu\nu}+\Gamma^{\omega}_{\mu\rho}\Gamma^{\sigma}_{\omega\nu}
-\Gamma^{\omega}_{\mu\nu}\Gamma^{\sigma}_{\omega\rho}$ for the covariant derivative and the Riemann tensor, respectively. The general form of the static cylindrical symmetric metric in the cylindrical weyl coordinates $(t,r,\varphi,z)$ is given by \cite{10de0810.4673}
\begin{eqnarray}
ds^2=g_{\mu\nu}dx^\mu dx^\nu=e^{2u(r)}dt^2-e^{2k(r)-2u(r)}\left(dr^2+dz^2\right)-w(r)^2e^{-2u(r)}d\varphi^2 \,\,.\label{ratbay3}
\end{eqnarray}

The non-null Christofell symbol read
\begin{eqnarray}
\Gamma^{t}_{tr}=u'\,,\,\Gamma^{r}_{tt}=u'e^{4u-2k}\,,\,\Gamma^{r}_{rr}=\Gamma^{z}_{rz}=-\Gamma^{r}_{zz}=k'-u'\,,  \Gamma^{r}_{\varphi\varphi}=-\left(ww'-u'w^2\right)e^{-2k}\,,\,\Gamma^{\varphi}_{r\varphi}=\frac{w'-u'w}{w}\,,
\end{eqnarray}
the non-null components of the Riemann tensor are
\begin{eqnarray}
R_{trrt}=\left(u''+2u'^2-k'u'\right)e^{2u}\,,\, R_{t\varphi\varphi t}=\left(u'ww'-u'^2w^2\right)e^{2u-2k}\,,\, R_{tzzt}=\left(k'u'-u'^2\right)e^{2u}\,,\, R_{rzzr}=\left(u''-k''\right)e^{2k-2u}\,,\nonumber\\
R_{r\varphi\varphi r}=-\Big[ww''-(u'+k')ww'+(k'u'-u'')w^2\Big]e^{-2u}\,,\,R_{\varphi zz\varphi}=\Big[\left(u'-k'\right)ww'+\left(k'u'-u'^2\right)w^2\Big]e^{-2u}\,.
\end{eqnarray}
The curvature scalar and the Gauss-Bonnet invariant read, respectively
\begin{eqnarray}
R=2\left(\frac{w''}{w}-\frac{u'w'}{w}-u''+u'^2+k'' \right)e^{2u-2k}\,\,,\label{ratbay6}\\
G=\frac{8e^{4(u-k)}}{w}\Big[ k'u'w''-u'^2w''-2u'u''w'+k'u''w'-u'^3w'+3k'u'^2w'+k''u'w'-2k'^2u'w'\nonumber\\+w\left(3u'^2u''-2k'u'u''+2u'^4-4k'u'^3-k''u'^2+2k'^2u'^2\right)\Big]\label{ratbay7}\,.
\end{eqnarray}
By using the cylindrical metric (\ref{ratbay3}) the equation (\ref{ratbay2}), one gets the following four equations
\begin{eqnarray}
\frac{8e^{4(u-k)}}{w}\Big\{ \left[u'w'-k'w'-w\left(u'^2-k'u'\right)\right]f''_{G}+
\Big[u'w''-k'w''+u''w'-2k'u'w'-k''w'\nonumber\\
+2k'^2w'-w\left(2u'u''-k'u''+u'^3-3k'u'^2-k''u'+2k'^2u'\right)\Big]f'_{G}
-\Big[u'^2w''-k'u'w''+2u'u''w'\nonumber\\
-k'u''w'+u'^3w'-3k'u'^2w'-k''u'w'+2k'^2u'w'-w\Big(3u'^2u''-2k'u'u''+2u'^4-4k'u'^3-k''u'^2\nonumber\\
+2k'^2u'^2\Big)\Big]f_{G}\Big\}+\frac{e^{2(u-k)}}{w}\Big[2u'w'-w''+w\left(2u''-u'^2-k''
\right)\Big]-f=\kappa^2\rho\,\,,\label{ratbay8}\\
\frac{8e^{4(u-k)}}{w}\Big\{ 3\left[u'^2w'+k'u'w'+w\left(u'^3-k'u'^2\right)\right]f'_G
+\Big[u'^2w''-k'u'w''+2u'u''w'-ku''w'+u'^3w'\nonumber\\
-3k'u'^2w'-k''u'w'+2k'^2u'w'+w\Big(2k'u'u''-3u'^2u''-2u'^4+4k'u'^3+k''u'^2-k'^2u'^2\Big)
\Big]f_{G}\Big\}\nonumber\\
+\frac{e^{2(u-k)}}{w}\Big[k'w'-wu'^2\Big]+f=\kappa^2 p_{r}\,\,,\label{ratbay9}
\end{eqnarray}
\begin{eqnarray}
\frac{8e^{4(u-k)}}{w}\Big\{w\left(k'u'-u'^2\right)f''_{G}+w\Big[k'u''-2u'u''-3u'^3+5k'u'^2+k''u'-2k'^2u'\Big]f'_{G}\nonumber\\
+\Big[u'^2w''-k'u'w''+2u'u''w'-k'u''w'+u'^3w'-3k'u'^2w'-k''u'w'+2k'^2u'w'
+w\Big(2k'u'u''\nonumber\\-3u'^2u''-2u'^4+4k'u'^3+k''u'^2-2k'^2u'^2\Big)\Big]f_{G}\Big\}
+e^{2(u-k)}\left(u'^2+k''\right)+f=\kappa^2 p_{\varphi}\,\,,\label{ratbay10}\\
\frac{8e^{4(u-k)}}{w}\Big\{ \left(u'w'-wu'^2\right)f''_G+\Big[u'w''+u''w'+2u'^2w-3k'u'w'
+w\Big(3k'u'^2-2u'u''-3u'^3\Big)\Big]f'_G\nonumber\\
+\Big[u'^2w''-k'u'w''+2u'u''w'-k'u''w'+u'^3w'-3k'u'^2w'-k''u'w'+2k'^2u'w'+w\Big(2k'u'u''\nonumber\\
-3u'^2u''-2u'^4+4k'u'^3+k''u'^2-2k'^2u'^2\Big)\Big]f_G\Big\}+\frac{e^{2(u-k)}}{w}\left(w''-k'w'+wu'^2\right)+f=\kappa^2 p_z\,\,.\label{ratbay11}
\end{eqnarray}

\section{ A specific solutions for $f(G)=\alpha G^n$}

In this section we present a more general type of Levi-Civita (LC) solution for power-law expression of the algebraic function $f$, in the form $f(G)=\alpha G^n$, with $\alpha,n\in\Re$. This model of $f(G)$ is a viable cosmological model \cite{viable,cognola1,bamba1,myrzakulov1}. Also, this model is a member of the general class of the models without the spurious spin-2 ghosts \cite{tsujikawa}. Also, this model is the leading order term of a viable, stable model of the Gauss-Bonnet gravity \cite{stability} and for local astrophysics phenomena \cite{cognola2}. 
\par
As it is well known in GR, LC solution corresponds to vacuum, where equivalently the Ricci scalar vanishes. Here, we present solutions for both metric parameters $w(r)$, $u(r)$ and $k(r)$, and also the the energy density and pressures of the matter content, where the curvature scalar is still  null. 
\par
To illustrate this we consider the  metric parameters as 
\begin{eqnarray}
w(r)=r\,\,,\label{ratbay12}\\
u(r)=u_0\ln{\left[w(r)\right]}\,\,,\label{ratbay13}\\
k(r)=k_0\ln{\left[w(r)\right]}\,\,,\label{ratbay14}
\end{eqnarray}
where $u_0$ and $k_0$ are constants. We must clarify this particular solution, specially we must know how we fix the gauge metric function $w(r)=r$. This last gauge is a harmonic function on the subspace $\Sigma_2$ defined by the two constant time and space like sheets $t=constant,\ \ r=constant$. It means it satisfy the potential equation $\nabla_A\nabla^A w(r)=0,\ \ A,B=\{z,\varphi\}$. As we know the solution of this  equation is a linear function of the radial coordinate $r$, as $w(r)=c_0r+c_1$. The constant can be set to the zero, $c_1=0$ by a shift symmetry of the conical cylinder geometry and the $c_0$ can be absorbed in the redefinition of the radial coordinate $r$. This last parameter induces a conical parameter $\hat{\alpha}=c_0$ which is related to the mass per length function \cite{aryal}. But for the pair of the metric functions $\{u(r),k(r)\}$, we follow the LC metric form in which these functions have been reinterpreted as the potential functions (purely Newtonian)of a rod with mass per length $M$ located at the center $r=0$. Also these functions are planar harmonic functions which satisfy the same potential equations, like the $w(r)$. These potential functions remain singular of the points near the axis $r=0$. The real thin cosmic string, which is an interior solution to LC metric must have a thickness with scale $r_0$. This thickness can not be found by the classical approaches. The essence of it backs to the quantum features of the naked singularities. Also here, the description of the cosmic string line element with the higher order corrections of the Gauss-Bonnet term is enable to solve the axis singularity problem. But also in this higher order corrected model, this assumption as an ansatz satisfy all the necessary requirements. This is verified that (\ref{ratbay12})-(\ref{ratbay14}) to be an exact solution to the vacuum Einstein field equations. With the choice  (\ref{ratbay12})-(\ref{ratbay14}) the curvature scalar (\ref{ratbay6}) and the Gauss-Bonnet invariant
(\ref{ratbay7}) read
\begin{eqnarray}
R(r)&=&-2\left(k_0-u_0^2\right)r^{2(u_0-k_0-1)}\label{ratbay15}\,\,,\\
G(r)&=&16u_0\left(u_0-1\right)\left(k_0-u_0\right)\left(k_0-u_0+1\right)r^{4\left(u_0-k_0+1\right)}\label{ratbay16}\,\,,
\end{eqnarray}
where the constant $u_0$ is now different from $1$.  Also, we get the energy density and pressures as follows
\begin{eqnarray}
\rho(r)=\frac{1}{\kappa^2}\Bigg\{(k_0-u_0^2)r^{2(u_0-k_0-1)}+\frac{\alpha 16^n(n-1)}{u_0}\Big[r^{4(u_0-k_0-1)}u_0(k_0-u_0)(1-k_0-u_0)(u_0-1)\Big]^n\nonumber\\
\times\Big[8n^2(1+k_0-u_0)+u_0+2n(3u_0-2k_0-2)\Big]\Bigg\}\label{ratbay17}\,\,,
\end{eqnarray}
\begin{eqnarray}
p_r(r)=\frac{1}{\kappa^2}\Bigg\{r^{2(u_0-k_0-1)}\left(k_0-u_0^2\right)+\frac{\alpha16^n(n-1)}{(k_0-u_0)(u_0-1)}\Big[r^{4(u_0-k_0-1)}u_0(k_0-u_0)(1+k_0-u_0)(u_0-1)\Big]^n\nonumber\\
 \times \Big[k_0(6n-1)(u_0-1)+u_0\left[u_0-1-6n(u_0+1)\right]\Big]\Bigg\}\label{ratbay18}\,\,,
\end{eqnarray}
\begin{eqnarray}
p_{\varphi}=\frac{1}{\kappa^2}\Bigg\{r^{2(u_0-k_0-1)}(u_0^2-k_0)+\frac{\alpha 16^n(n-1)}{u_0-1}\Big[r^{4(u_0-k_0-1)}u_0(k_0-u_0)(1+k_0-u_0)(u_0-1)\Big]^n\nonumber\\
 \times \Big[1-u_0+2n(u_0-2k_0-1)+8n^2(1+k_0-u_0)\Big]\Bigg\}\label{ratbay19}\,\,,
\end{eqnarray}
\begin{eqnarray}
p_z(r)&=&-\frac{1}{\kappa^2}\Bigg\{r^{2(u_0-k_0-1)}(k_0-u_0^2)+\frac{\alpha 16^n(n-1)}{(k_0-u_0)(u_0-1)}\Big[r^{4(u_0-k_0-1)}u_0(k_0-u_0)(1+k_0-u_0)(u_0-1)\Big]^n\nonumber\\
 &\times& \Big[k_0(1-2n+8n^2)(u_0-1)+u_0(1-u_0)-8n^2(u_0-1)^2+2n\left[2+u_0(u_0+2r-5)\right]\Big]\Bigg\}\label{ratbay20}\,\,.
\end{eqnarray}

Since our goal is to get a new type of LC solution \cite{LC}, we need a null curvature scalar. We want to label the formation of the LC family to the effects of the Gauss-Bonnet terms in the action. It means we want to know how we can recover LC solution in the absence of the curvature scalar by including the Gauss-Bonnet invariants. The only way for this is to set $k_0=M^2$ and $u_0=M$. This set of the parameters came from the constraint $k_0=u_0^2$. This choice of the parameters $\{u_0,k_0\}$ parametrize the family of the solutions as a one parameter family of the exact solution with the singly parameter of the mass per length $M$. There is no hair from the Gauss-Bonnet coupling $\alpha$. Therefore, the Gauss-Bonnet invariant becomes
\begin{eqnarray}
G(r)=16M^2\left(M-1\right)^2\left(M^2-M+1\right)r^{-4\left(1-M+M^2\right)}\label{ratbay21}\,\,.
\end{eqnarray}
In the same way, the energy density and the pressures can be found as
\begin{eqnarray}
\rho(r)&=&\frac{\alpha}{\kappa^2}16^n\frac{(n-1)M^{2n-1}}{r^{4n(M^2-M+1)}}(M-1)^{2n}(M^2-M+1)^n\Big[M-2n(2M^2-3M+2)+8n^2(M^2-M+1)\Big]\label{ratbay22}\,\,,
\end{eqnarray}
\begin{eqnarray}
p_r(r)=\frac{M\left[6n(M^2-2M-1)-(M-1)^2\right]}{(M-1)^2\left[M-2n(2M^2-3M+2)+8n^2(M^2-M+1)\right]}\rho(r)\label{ratbay23}\,\,,
\end{eqnarray}
\begin{eqnarray}
p_\varphi(r)=\frac{M\left[1-M-2n(2M^2-M+1)+8n^2(M^2-M+1)\right]}{(M-1)\left[M-2n(2M^2-3M+2)+8n^2(M^2-M+1)\right]}\rho(r)\label{ratbay24}\,\,,
\end{eqnarray}
\begin{eqnarray}
p_z(r)=\frac{\left\{M^2-M+8n^2(M-1)^2-M^2(M-1)(8n^2-2n+1)-2n\left[M^2+M(2r-5)+2\right]\right\}}{(M-1)^2\left[M-2n(2M^2-3M+2)+8n^2(M^2-M+1)\right]}\rho(r)\label{ratbay25}\,\,.
\end{eqnarray}

For we establish the positivity of $\rho(r)$ we must have the following satisfied conditions. For $\alpha>0$, $n\leq 0$ and $M>0$, or $0<n<1/2$ and $M_1\leq M\leq M_2$, or $1/2 \leq n\leq 1$ and $M>0$. For $\alpha<0$, $0<n<1/2$ and $0<M\leq M_1$ or $M_2\leq M$, or still $1\leq n$ and $M>0$. We have $M_{1,2}=[(8n^2-6n-1)/8n(2n-1)]\pm (1/8)\sqrt{(1+12n-44n^2+
166n^3-192n^4)/n^2(2n-1)^2}$.  
\par
We also noticed that we have a behavior of the type barotropic for the components of pressure. Two are homogeneous , $p_r=\omega_r \rho$ and $p_{\phi}=\omega_{\phi}\rho$, and one inhomogeneous, $p_z=\omega_z (r)\rho$.
\par
Note that for $\alpha=0$ the algebraic function $f(G)$, the energy density and pressures vanish, leading to the usual LC solution solution in GR, which indeed is a vacuum solution. We see here that the type of LC solution in power-law modified $f(G)=\alpha G^n$ gravity is not a vacuum solution within the assumption (\ref{ratbay12})-(\ref{ratbay14}), but falls into a vacuum solution for $\alpha=0$ (the usual LC solution in GR). 
\par
Other interesting feature to  be put out here is the case where the parameter $n=1$ and $\alpha \neq 0$. Observe that in this case, the algebraic function $f$ has a linear form in $G$, that is $f(G)=\alpha G$, which is a topological term in the action. Within this case, one sees that the $G$ contribution fades away from the field equations leading directly to the LC  solution, once the conditions $k_0=M^2$ and $u_0=M$ are satisfied, where the curvature scalar is null. We mention here that it is possible to Match this static solution to the exterior metric of a cosmic string \cite{vilenkin,linet,tian}. Also a rotating version of this solution can be matched successfully to the Kerr solution \cite{Kyriakopoulos}.

\section{Conclusion}
Cosmic strings are the exact solutions of the classical Einstein-Hilbert action with cylindrical symmetry. They generated in the early universe and in the high energy regime as the products of the spontaneous symmetry breaking of the $U(1)$ abelian gauge fields coupled non minimally to the Higgs bosons. The topology of this flat solutions is non trivial and so this different type of the topology differers them from the Minkowski spacetimes. In this paper we investigated the possibility to have a cylindrically symmetric vacuum solution in a modified Einstein gravity with Gauss-Bonnet higher order corrections. We derived the general field equations which they have been described the general vacuum solution of the Levi-Civita with a viable Gauss-Bonnet model  $f(G)\sim G^n,n\neq1$. We showed that the system of the gravitational field equations recover a modified version of the LC solutions as a special case with some analytic functions of the fluid's energy momentum components. Explicitly we observed that the solution satisfies the field equations properly by direct replacement. Our solution can be considered as a generalization of the exterior solution of a Cosmic string in the modified Gauss-Bonnet gravity.

\vspace{0.5cm}
{\bf Acknowledgement:} M. E. Rodrigues thanks a lot UFES for the hospitality during the elaboration of this work and  M. J. S. Houndjo  thanks CNPq/FAPES for financial support.

\end{document}